# Long lifetime and high-fidelity quantum memory of photonic polarization qubit by lifting Zeeman degeneracy

*Zhongxiao Xu, Yuelong Wu, Long Tian, Lirong Chen, Zhiying Zhang, Zhihui Yan, Shujing Li, Hai Wang[*], Changde Xie, Kunchi Peng*

*The State Key Laboratory of Quantum Optics and Quantum Optics Devices, Institute of Opto-Electronics, Shanxi University, Taiyuan, 030006, People's Republic of China*

Long-lived and high-fidelity memory for photonic polarization qubit (PPQ) is crucial for constructing quantum networks. Here we present an EIT-based millisecond storage system in which a moderate magnetic field is applied on a cold-atom cloud to lift Zeeman degeneracy. PPQ states are stored as two magnetic-field-insensitive spin waves. Especially, the influence of magnetic-field-sensitive spin waves on the storage performances is almost totally avoided. The measured average fidelities of polarization states are 98.6% at 200 μs and 78.4% at 4.5 ms, respectively.

Photonic polarization qubits (PPQs) are extensively used for encoding and transmitting quantum information since they are easily manipulated and analyzed [1-6]. For realizing remote quantum communication [1, 7-8] and distributed quantum computation [1, 9] we have to achieve the faithful storage and retrieving of PPQs. A variety of physical processes, such as electromagnetically induced transparency (EIT) [10-12], spontaneous Raman emission (SRE) [3, 13-14], atomic-frequency combs [15-16], off-resonant Faraday interaction [17], far-off-resonant Raman interaction [18] and Gradient echo [19], have been exploited to store quantum states of light. The dynamic EIT and SRE processes in cold atoms provide promising storage schemes [1, 7]. The SRE process is an elementary step in DLCZ protocol [7-8], which is suitable to generate a heralded entanglement between two long-distance atomic ensembles [10]. However, the low probabilistic success of the scheme in preparing such entanglement limits its application in quantum information [10,12]. To overcome this drawback, alternative approaches based on separating the two processes for the generation and storage of qubits have been proposed [10-12, 20-21]. A typical way is to prepare two polarization-entangled photons firstly, then transport them into two remote nodes and store them in the nodes [21], respectively. This approach promises to generate a deterministic polarization-entanglement between two remote memories [21]. The dynamic EIT is a promising process for realizing the storage of single-photon polarization qubits since it can directly receive and preserve the quantum states of photons coming from outside systems while the SRE process is not able to do so [1]. Unlike general storages for a certain polarization state of light which only require a single spin wave (SW) and have been experimentally realized with high retrieval efficiency up to ~78% [22] and long lifetime up to ~0.2 s [23-24]. For the storage of a PPQ, we must store its logical states as a superposition of two SWs [3]. Toward realizing the quantum repeater, several PPQ storage experiments have been implemented in quantum region [10-12]. In the experiments of Ref.[10-11], single-photon polarization qubits are split into vertical- and horizontal-components and then stored in two atomic systems respectively placed at two space-separated arms of an interferometer, and the achieved storage lifetime is several microseconds. Another EIT-based storage experiment of PPQ is realized in a Bose-Einstein condensation (BEC), in which the memory qubit is preserved in two atomic magnetic-field-sensitive coherences $m=0 \leftrightarrow m'=\pm 1$ and the residual magnetic field is actively compensated to reduce the decoherence [12]. The measured polarization fidelity is ~0.95 for the storage period of 2 μs and ~0.75 for 470 μs, respectively, which is the longest lifetime of the qubit memory realized with EIT-based scheme, so far. To increase the storage lifetime PPQs should be encoded in atomic coherences associated with magnetic-field-insensitive transitions, which have been utilized in the DLCZ-type experiments [13-14]. In Ref.[13], Kuzmich's group creates a memory qubit whose logical states are preserved in two SWs associated with the magnetic-field-insensitive coherences $m = \pm 1 \leftrightarrow m' = \mp 1$ of two overlapped atomic ensembles, which is confined in a 1D optical lattice. Although the experiment demonstrates the violation of Bell's equality for storage time of 3 ms, the unwanted magnetic-field-sensitive SWs are also produced during the creation of the qubit memory, which leads the retrieval efficiency to decrease promptly with the storage time [25]. That is because the magnetically-sensitive coherences are washed out fast within 100 μs [13]. Besides, due to the interference between the clock SW ($m = 0 \leftrightarrow m' = 0$) and the magnetic-field-insensitive SWs, the higher entanglement appears only at the periodical interval $T = 0.54n\ ms,\ (n = 1,2,...6)$ [13]. Also, the same group realizes the quantum memory with a lifetime of 100 ms by encoding qubit states in two spatially-distinct SWs associated with the $m = 0 \leftrightarrow m' = 0$ clock coherence and applying a magic-valued magnetic field to eliminate the lattice-induced dephasing [14]. However, because of the spatially splitting of the SW qubit states, some extra requirements, such as the interferometric stability [7] and spatially matching of the two SW modes, have to be added.

Until now, the EIT-based scheme of storing PPQs as magnetic-field-insensitive SWs has not been implemented. Here, we present an effective long lifetime and high-fidelity EIT-based storage experiment for PPQs. By applying a moderate magnetic field on a cold [87]Rb atomic cloud, only two pairs of magnetic-field-insensitive transitions appear respectively in two EIT systems existing in a cold atomic cloud, which will be used for storing the PPQ states. At the same time, all magnetic-field-sensitive transitions are removed outside the EIT systems when the degeneracy of Zeeman sublevels is lifted. Thus, the influences of magnetic-field-sensitive coherences on the storage are eliminated and the performances of the qubit memory are significantly improved.

The involved levels of $^{87}$Rb atoms are shown in Fig.1(a) and (b), where $|a\rangle = |5^2S_{1/2}, F=1\rangle$, $|b\rangle = |5^2S_{1/2}, F=2\rangle$, $|e\rangle = |5^2P_{1/2}, F'=1\rangle$, respectively. By optical pumping, half of the cold atoms can be prepared in state $|a_{m=1}\rangle$ and other half in state $|a_{m=-1}\rangle$ ($m$ denotes the magnetic-quantum number), thus the cloud of cold atoms is composed of two incoherent spatially-overlapped atomic ensembles. The frequencies of the signal and writing/reading light fields are tuned to transitions $|a\rangle \leftrightarrow |e\rangle$ and $|b\rangle \leftrightarrow |e\rangle$, respectively, their frequency difference is $\omega_{ab}$, which matches the resonance frequency of the two-photon transition $|a\rangle \leftrightarrow |b\rangle$ at the case of Zeeman degeneracy. For suppressing the dephasing effect resulting from atomic motion, we make the signal and writing/reading light beams collinearly go through the cold-atom cloud along z-direction. Such collinear configuration has been first proposed and demonstrated by Zhao et. al. [26] in the DLCZ-type experiment, in which they have achieved the storage lifetime of ~1 ms for single photons with a fixed-polarization. In the presented experiment, the input signal-light

field may be set in an arbitrary polarization state, which can be regarded as the superposition of the right ($\sigma^+$) and left ($\sigma^-$) circular polarization components. Since the quantum axis is defined by applying a bias magnetic field $B_0$ along z-direction, the $\sigma^+$- and $\sigma^-$-polarized components of the signal-light field couple to $|a_m\rangle \leftrightarrow |e_{m+1}\rangle$ and $|a_m\rangle \leftrightarrow |e_{m-1}\rangle$ transitions, respectively. The writing/reading light field is vertically-polarized, its right- and left-circular-polarized components ($W^\pm / R^\pm$) drive $|b_m\rangle \leftrightarrow |e_{m+1}\rangle$ and $|b_m\rangle \leftrightarrow |e_{m-1}\rangle$ transitions, respectively. In previous EIT-based storages of PPQs [10, 12], the typical value of the magnetic fields used to define quantization axis is about several hundreds milliGauss. When such a weak field is applied on the $^{87}$Rb atomic ensembles, the degeneracy of the Zeeman sublevels of the $F=1$ and $2$ ground states can't be lifted (see Fig.1(a)). In this case, for the $\sigma^+$- ($\sigma^-$-) polarized component, the EIT occurs in the four-level tripod system [27] formed by $|a_{m=-1}\rangle - |b_{m=-1}\rangle - |e_{m=0}\rangle - |b_{m=1}\rangle$ ($|a_{m=1}\rangle - |b_{m=1}\rangle - |e_{m=0}\rangle - |b_{m=-1}\rangle$). By switching off the writing beam, $\sigma^+$- ($\sigma^-$-) polarized component of the input signal is transferred into the SWs $S_{-1,1}$ ($S_{1,-1}$) and $S_{-1,-1}$ ($S_{1,1}$), and stored in the cloud of cold atoms, where $S_{-1,1}$ ($S_{1,-1}$) is associated with the magnetic-field-insensitive coherence $|a_{m=-1}\rangle \leftrightarrow |b_{m=1}\rangle$ ($|a_{m=1}\rangle \leftrightarrow |b_{m=-1}\rangle$), while $S_{1,1}$ ($S_{-1,-1}$) is associated with the magnetic-field-sensitive coherence $|a_{m=1}\rangle \leftrightarrow |b_{m=1}\rangle$ ($|a_{m=-1}\rangle \leftrightarrow |b_{m=-1}\rangle$). In our experiment, the degeneracy of Zeeman sublevels is obviously lifted (see Fig.1(b)) by applying a moderate magnetic field along z-direction. The frequency of the magnetic-field-insensitive $|a_{m=1}\rangle \leftrightarrow |b_{m=-1}\rangle$ ($|a_{m=-1}\rangle \leftrightarrow |b_{m=1}\rangle$) transition still matches $\omega_{ab}$, while the frequency of the magnetic-field-sensitive $|a_{m=1}\rangle \leftrightarrow |b_{m=1}\rangle$ ($|a_{m=-1}\rangle \leftrightarrow |b_{m=-1}\rangle$) transition becomes mismatching $\omega_{ab}$. Therefore, the four-level tripod EIT system will change to three-level $\Lambda$-type EIT system formed by $|a_{m=-1}\rangle - |e_{m=0}\rangle - |b_{m=1}\rangle$ ($|a_{m=1}\rangle - |e_{m=0}\rangle - |b_{m=-1}\rangle$) and the $\sigma^+$- or $\sigma^-$-polarized component of signal photons will be only transferred into the magnetic-field-insensitive SW $S_{-1,1}$ or $S_{1,-1}$. By using dark-state polariton concepts [27-29], we derive the retrieval efficiencies of the $\sigma^+$- or $\sigma^-$-polarized signal photons, which are

$$R_e^\pm(t) = R_{e0} e^{-\frac{t}{\tau_1}}, \qquad (1)$$

for the case of the lifting-Zeeman degeneracy, and

$$R_e^\pm(t) = \frac{1}{4} R_{e0} \left| e^{\frac{-t}{2\tau_1} - i\omega_{\mp1,\pm1}t} + e^{\frac{-t}{2\tau_{\mp1,\mp1}} - i\omega_{\mp1,\mp1}t} \right|^2 \qquad (2)$$

for the case of the Zeeman degeneracy (See supplemental material [30] for details), respectively.

Before the experiments, we theoretically evaluate the retrieval efficiencies for both cases of lifting and no-lifting Zeeman degeneracy respectively through Eqs.(1) and (2). In the evaluation, we take $\tau_1 = 1ms$ [26] and $\tau_2 = 50\mu s$ [13]. We find that the retrieval efficiency of the $\sigma^+$- ($\sigma^-$-) polarized signal photons for the lifting-Zeeman-degenerate case is about 4 times that for the Zeeman-degenerate case at times longer than ~100 µs. The physical reason is that the partial optical signals are transferred into the SW $S_{1,1}$ ($S_{-1,-1}$) with fast decay rate for the Zeeman-degenerate case. Following, we perform the storage and retrieval experiments of $\sigma^+$- ($\sigma^-$-) polarized signal light with the input peak power of 25 µW for the two cases to confirm the theoretical expectation.

The experimental setup for the storage of PPQ is shown in Fig.1(c). A cold atomic cloud including about ~$10^9$ $^{87}$Rb atoms serves as the quantum memory medium. By using the lasers P1, P2, and P3 (for details see Supplemental Material [30]) to optically pump it, half of the cold atoms are prepared in state $|a_{m=1}\rangle$ and other half in state $|a_{m=-1}\rangle$. The input signal and writing/reading light beams are combined by using a polarization-insensitive beam-splitter BS. Before arriving BS, the signal beam goes through neutral density filters (NDFs), a quarter-wave plate (QWP1) and a half-wave plate (HWP1) as well as the writing/reading light beam passes through two optical mode cleaners MC$_{1, 2}$, respectively. With QWP1 and HWP1, the polarization state of the signal light can be arbitrarily set. The optical mode cleaners MC$_{1,2}$ are used to filter the incoherent components of the reading laser pulses (for details see Supplemental Material [30]). After BS, the signal and the writing/reading beams collinearly propagate through the cold atoms along z-direction. The diameters of the signal and writing/reading beams in the center of cold atoms are ~1 mm and ~1.4 mm. The power of the writing/reading beam is 1.3 mW. The retrieval signal photons and writing/reading beam go through a pinhole (PH) and optical spectral filters OSF (for details see Supplemental Material [30]). With PH and OSF, the stray light and writing/reading beam can be blocked. After OSF, the retrieval photons go through a half-wave plate HWP$_2$ and then enter into a polarization analyzing and measuring (PAM) system. HWP$_2$ is used to add a phase shift $\Phi = -\phi$ between the $\sigma^-$- and $\sigma^+$-polarized components of the retrieval photons to compensate the relative phase $\phi$ between the SWs $S_{1,-1}$ and $S_{-1,1}$, where $\phi = (\omega_{1,-1} - \omega_{-1,1})t$, which is induced by atomic Larmor precession, $\omega_{1,-1}$ ($\omega_{-1,1}$) is the Larmor frequency of $S_{1,-1}$ ($S_{-1,1}$) at a dc magnetic field $B_0$, and $t$ is storage time. PAM consists of a quarter-wave plate (QWP2) a half-wave plate (HWP3), a Glan-laser polarizer (GLP), detectors D$_1$ and D$_2$. With QWP2 and HWP3, we can select the polarization basis $H-V$, $L-R$ or $D-A$ in turn for the polarization analyzing and measuring, where $H$, $V$, $R$, $L$, $D$ and $A$ denote horizontal, vertical, right circular, left circular, diagonal (45°), antidiagonal (-45°) polarization, respectively. The output photons from PBS's two ports are respectively coupled into two fibers and then are detected by D$_1$ and D$_2$, respectively. D$_1$ and D$_2$ are photodiode detectors in the measurement of retrieval efficiencies in Fig.2 and single-photon detectors in polarization fidelity measurements in Fig.3 and Table 1, respectively.

In each experimental trial, the cold $^{87}$Rb atomic cloud with the temperature of about 200 μK is collected by the magneto-optical trap (MOT) within ~520 ms. Then, trapping magnetic field and repumping laser are switched off and the detuning (with respect to the transition $\left|5^2S_{1/2}, F=2\right\rangle \leftrightarrow \left|5^2P_{3/2}, F'=3\right\rangle$) of the cooling laser is changed from -24.5 MHz to -39.5 MHz for Sisyphus cooling. After 0.7 ms, the cooling laser is turned off and the bias magnetic field is switched on. Waiting for 0.5 ms to make the bias field reach to a stabilization value (0.59 G or 13.5 G), then the pumping lasers P1, 2, 3 and the writing coupling laser are turned on. Keeping the optical pumping for 18 μs, we estimate that the most of atoms have been prepared into the states $\left|a_{m=1}\right\rangle$ or $\left|a_{m=-1}\right\rangle$ with equal populations, and the measured optical depth of the cold atoms for the transition $\left|a_{m=\pm1}\right\rangle \leftrightarrow \left|e_{m=0}\right\rangle$ is ~4. After the optical pumping, i.e. at the time $t$=0, the $\sigma^+$-polarized signal pulses (with a pulse length of 100 ns) are switched on. At the falling edge of the signal pulse, the writing laser beam is ramped to zero and thus the signal input pulse is stored into two atomic ensembles respectively. After a variable time delay t, we switch on the reading beam to retrieve the stored spin waves and detect the retrieved photons within a window of ~100 ns. The total duration for a measurement trial is ~50 ms, after which the measurement interval is terminated and a new MOT is prepared for next trial.

The measured retrieval efficiencies of $\sigma^+$- and $\sigma^-$-polarized signal fields as the function of storage time at $B_0$=0.59 G are shown in Fig.2(a) and (b) (red circle dots), respectively. We can see that they fast drop to very low levels around ~80 μs, that is because the magnetic-field-sensitive SWs are washed out at times longer than ~80 μs. The solid green lines II and II' are the fits to the experimental data based on theoretical models. In the models, the influences of imperfect atomic preparation have been taken into account (see supplemental material [30]), and then the fittings are further in agreement with the measured results. The black square dots in Fig.2(a) and (b) are the measured retrieval efficiencies of $\sigma^+$- and $\sigma^-$-polarized signal fields at $B_0$=13.5 G, respectively, which can be fitted by using Eq.(1) and the fittings yield a storage lifetime value $\tau_1 = 1.6$ $ms$ and the retrieval efficiency $R_{e0} = 8.3\%$ at $t$=0. Comparing the curves I (I') and II (II'), we find that the measured retrieval efficiencies at $B_0$=13.5 G is ~4 times that at $B_0$=0.59 G at times longer than ~80μs, which is in agreement with our theoretical prediction. At $B_0$=13.5G, the two retrieval efficiencies for $\sigma^\pm$-polarized signal photons are symmetric and the storages are long-lived, which promise us to achieve a high-fidelity and long lifetime storage of PPQ.

Subsequently, we perform the storage and retrieval of PPQ at single-photon level for the dc magnetic field $B_0$=13.5 G. The input signal pulse is decreased to single-photon level (i.e., the mean photon number $\bar{n}=1$) by the neutral density filters (NDF). We determine the mean photon number by measuring the detection probability per pulse when the cold atomic cloud does not exist, in which, the total detection efficiency of $\eta_d \approx 19\%$ is taken into account (See supplemental material [30]).

To characterize the quality of the qubit memory, we perform the experiments of the storage and retrieval for four input polarization states $|H\rangle$, $|V\rangle$, $|D\rangle$ and $|R\rangle$, respectively. By analyzing the retrieval photon in three orthogonal bases $|H\rangle-|V\rangle$, $|R\rangle-|L\rangle$ and $|D\rangle-|A\rangle$, we reconstruct the density matrix $\rho_{out}$ of the retrieved single photons by means of the quantum state tomography [31]. The fidelity of the quantum state is defined as the overlap of the density matrix $\rho_{out}$ with the ideal input state $|\psi_i\rangle$: $F_{st}=\langle\psi_i|\rho_{out}|\psi_i\rangle$. The fidelities of the four input states at several different storage times are listed in Table 1. The measured average fidelity is 98.6% at 200 μs and decreases to 78.4% at 4.5 ms.

Alternatively, the storage of PPQ can be characterized by the quantum process matrix $\chi$ [32]. After reconstructing the matrix $\chi$ (see Supplemental Material [30]), we obtain $F_{process}$. The function of $F_{process}$ as the storage time is shown in Fig.3, we can see that $F_{process}$ decreases with the storage time. We attribute such decrease to the following two factors. The retrieval efficiency exponentially reduces with the storage time, which makes the background noise gradually become a main contribution to the single-photon-counting events and thus the polarization fidelity is degraded. On the other hand, the dephasing between the two spin waves $S_{1,-1}$ and $S_{-1,1}$ induced by the temporal fluctuations of the magnetic field in z-direction will reduce the retrieval fidelity also. Considering all above-mentioned factors, a formula used for evaluating the quantum process fidelity is deduced (see Supplemental Material [30]):

$$F_{process} \approx \frac{(1+\gamma(t))\eta_d R_e(t) + N}{2(\eta_d R_e(t) + 2N)} \quad (3)$$

where $\gamma(t) = \exp[-t^2/\sigma_\gamma^2]$ is the dephasing factor. $\eta_d$ is the total detection efficiency, $R_e(t)$ is the retrieval efficiency. $N$ corresponds to the relative background noise. The solid line is the fitting to the data of the quantum-process fidelity according to the formula (3), which is well in agreement with the measured results. The fitting yields a $e^{-1}$ dephasing time of $\sigma_\gamma = 14\ ms$, which corresponds to the magnetic-field temporal fluctuations of $\sigma_B \approx 3\ mG$ (see supplemental Material [30]). If an active compensation technology is utilized to reduce the magnetic-field fluctuations to $\sigma_B \approx 0.3\ mG$, the dephasing time of ~100 ms is expected. Besides, after the writing process finishes, if reducing the magnetic field from ~13.5 G to 3.23 G during storage, the used coherences $|a_{m=\pm 1}\rangle \leftrightarrow |b_{m=\mp 1}\rangle$ will become the perfect first-order magnetically insensitive [23], and thus the dephasing time will further increase.

In summary, we have proposed and experimentally demonstrated an effective EIT-based approach to achieve the long lifetime and high-fidelity storage for PPQs. By means of lifting Zeeman degeneracy and removing the magnetic-field-sensitive sublevels out of EIT systems, the input signal photons are only mapped on two magnetic-field-insensitive SWs, and the bad influences of magnetic-field-sensitive SWs to the storage ability are eliminated. Thus, the storage performance is significantly improved. For the storage time of less than 4.5 ms, the average fidelity measured in the presented qubit memory is beyond 78%, which is the lowest boundary for the violation of the Bell inequality [34]. The demonstrated approach is robust due to no requirement of the interferometer stability between the two spatially-separated modes. The lifetime of the qubit memory in the presented system is mainly limited by the retrieval efficiency, the dephasing time between two spin waves and the background noise. If the atomic cloud is placed in an optical cavity [33], the retrieval efficiency can be further enhanced and the polarization fidelity can be further improved. We believed that the demonstrated qubit memory approach can be utilized to store the polarization-entangled photon pairs or single-photon polarization qubit [12] from trapped single atom for realizing long-distance quantum communication [1, 7, 21] and implementing distributed quantum computing [35].

*Corresponding author: wanghai@sxu.edu.cn

**Acknowledgement:**

We acknowledge funding support from the 973 Program (2010CB923103), the National Natural Science Foundation of China (No.10874106, 11274211, 60821004).

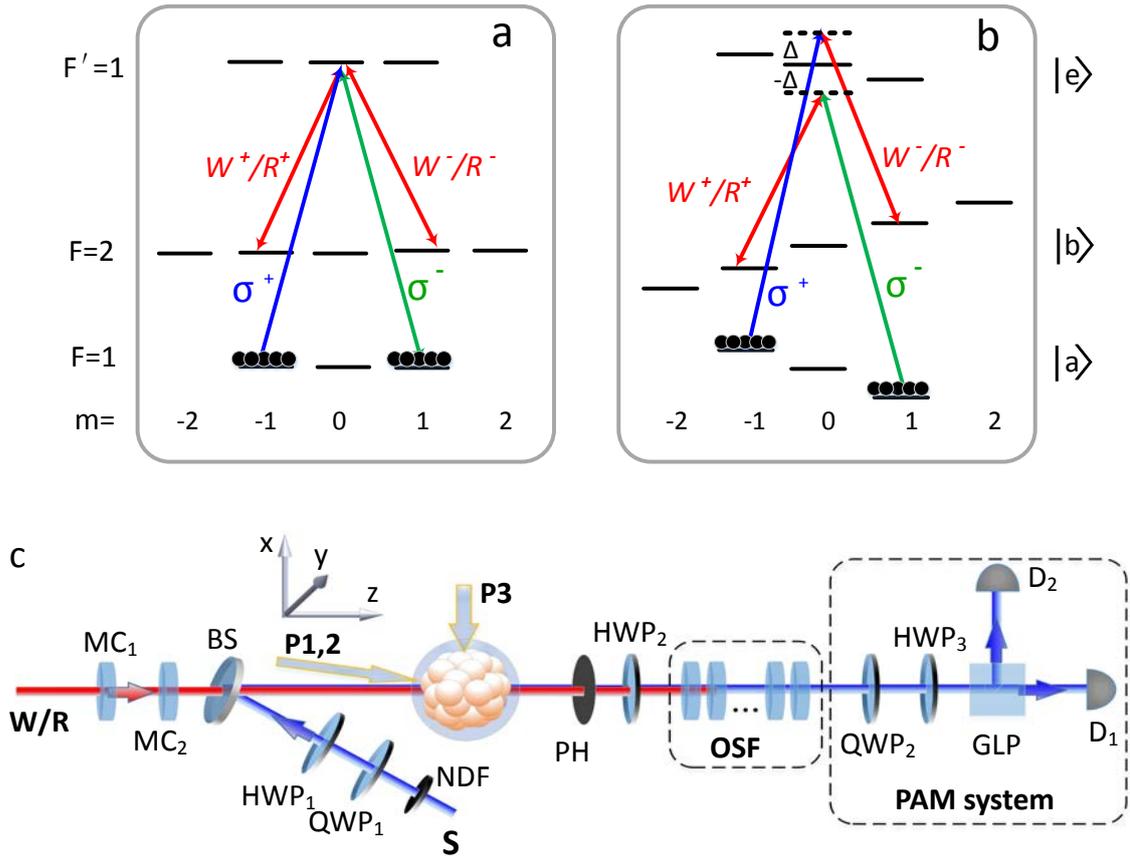

**Fig.1 Overview of the experiment.** (a) and (b) are the atomic level schemes of $^{87}$Rb with a weak ($B_0$=0.59 G) and a moderate ($B_0$=13.5 G) magnetic field in z-direction, respectively. $\sigma^+$ ($\sigma^-$) stands for the right- (left-) polarized input signal light field. $W^+/R^+$ and $W^-/R^-$ denote the right- and left-polarized writing/ reading fields, respectively. (c) The experimental set up. The polarization of the signal (S) beam can be set arbitrarily by a quarter-wave plate QWP1 and a half-wave plate HWP1. The signal and writing/reading (W/R) light beams are combined at a polarization-insensitive beam-splitter BS and then collinearly propagate through the cold $^{87}$Rb atoms. For $B_0$=13.5G, the polarization qubit states can be stored as two

magnetic-field-insensitive SWs and then retrieved by dynamic EIT. The retrieval photons are sent into a polarization analyzing and measuring (PAM) system.

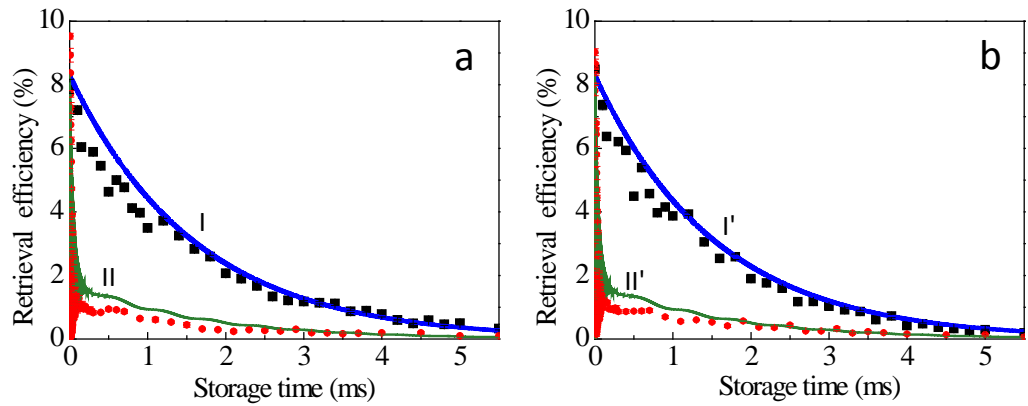

**Fig.2 Measured retrieval efficiencies ($R_e$) as the function of the storage time *t* for the input signal light with a peak power of ~25 μW.** (a) and (b) show the retrieval efficiencies of the $\sigma^+$- and $\sigma^-$-polarized input signal light, respectively. Black square and red circular points are the experimental data obtained for the lifting degenerate ($B_0$=13.5G) and degenerate ($B_0$=0.59G) cases, respectively. The blue solid lines in a and b are the fittings to the experimental data (black square points) according to $R_e(t) = R_{e0}e^{-t/\tau_1}$, respectively, which yields $R_{e0}$= 8.3%, $\tau_1 = 1.6\ ms$. The green lines in (a) and (b) are the fittings to the experimental data (red circular points) according to theoretical models in which the imperfect atomic preparation has been taken into account (see supplemental material [30]).

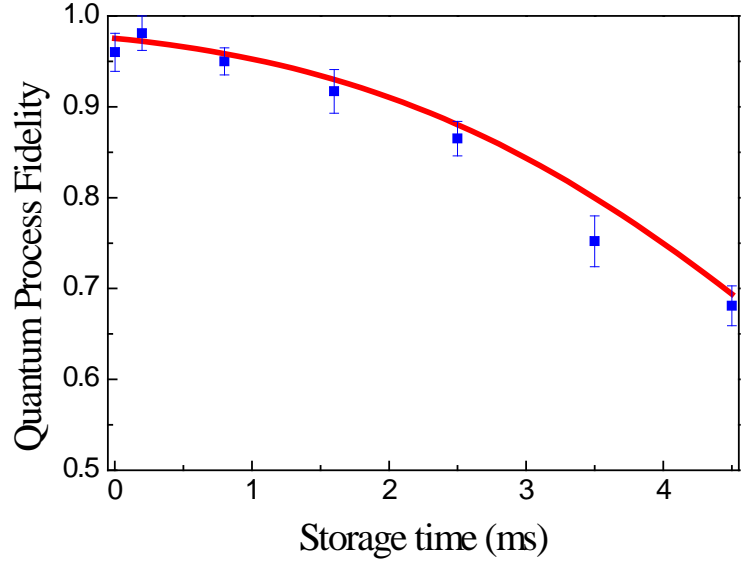

**Fig.3 The quantum process fidelity as the function of the storage time *t*.** The data are fitted by the expression $F_{process} \approx \dfrac{(1+e^{-t^2/\sigma_\gamma^2})\eta_d R_e(t)+N}{2(\eta_d R_e(t)+2N)}$ , where, the measured total detection efficiency $\eta_d \approx 0.19$, the measured background noise $N = 2.2\times 10^{-4}$. The retrieval efficiency $R_e(t) = 0.083 e^{-t/1.6}$ is obtained from both fitting curves I and I' in Fig.2. The fitting yields the dephasing time of $\sigma_\gamma = 14\ ms$.

**Table1 Quantum state fidelities of the four input polarization states for several storage times.** $F_{st(X)}$ are the measured state fidelities respectively for four different input polarized states of photons (X =H, V, D, R) without any background noise subtraction; $F_{ava} = \left(F_{st(H)} + F_{st(V)} + F_{st(D)} + F_{st(R)}\right)/4$ is the average fidelity; t is the storage time. The errors are obtained by Monte Carlo simulation which takes into account the statically uncertainty of photon counts.

| $t$(μs) | $F_{st(H)}$ (%) | $F_{st(V)}$ (%) | $F_{st(D)}$ (%) | $F_{st(R)}$ (%) | $F_{ava}$(%) |
|---|---|---|---|---|---|
| 2 | 96.7±1.2 | 98±1.1 | 97.7±1.1 | 98.7±0.85 | 97.8±1.06 |
| 200 | 98.3±1.2 | 98.1±1.1 | 98.1±1.1 | 99.8±0.19 | 98.6±0.89 |
| 800 | 95.5±1.5 | 97±1.3 | 96±1.3 | 97.6±1.2 | 96.5±1.33 |
| 1600 | 94.1±1.5 | 94.5±1.7 | 94.6±1.6 | 96.4±1.3 | 94.9±1.5 |
| 2500 | 90.7±2.0 | 91±2.0 | 89±2.0 | 91.9±2.0 | 90.7±2 |
| 3500 | 85.7±2.3 | 82.6±2.6 | 87.7±2.3 | 84±2.6 | 84.0±2.45 |
| 4500 | 82±2.8 | 72±3.2 | 79.4±2.9 | 80±2.9 | 78.4±2.95 |

# Supplemental Material

# Long lifetime and high-fidelity quantum memory of photonic polarization qubit by lifting Zeeman degeneracy

Zhongxiao Xu, Yuelong Wu, Long Tian, Lirong Chen, Zhiying Zhang,

Zhihui Yan, Shujing Li, *Hai Wang, Changde Xie, Kunchi Peng

The State Key Laboratory of Quantum Optics and Quantum Optics Devices,

Institute of Opto-Electronics, Shanxi University, Taiyuan, 030006,

People's Republic of China

## Experimental details

The $\sigma^-$-polarized pumping laser P1 and $\sigma^+$-polarized pumping laser P2 collinearly propagate through the atoms with a deviation angle ~2° from z-direction, which drives the transitions $\left|5^2S_{1/2}, F=2, m_F\right\rangle \leftrightarrow \left|5^2P_{1/2}, F'=2, m_F\text{-}1\right\rangle$ ($m_F$=2 to -1) and $\left|5^2S_{1/2}, F=2, m_F\right\rangle \leftrightarrow \left|5^2P_{1/2}, F'=1, m_F\text{+}1\right\rangle$ ($m_F$=-2 to 0), respectively. The π-polarization pumping laser P3 propagates through the atoms along x-direction, which drives the transition $\left|5S_{1/2}, F=1, m_F=0\right\rangle \leftrightarrow \left|5^2P_{3/2}, F'=0, m_F=0\right\rangle$. In the center of cold atoms, the powers and diameters of the lasers P1, P2 in the center of cold atoms are approximately equal, which are ~10 mW and ~7 mm, respectively, while that of the laser $P_3$ is ~400 µW and ~8.8 mm respectively.

The optical mode cleaners $MC_1$ and $MC_2$ are two ring optical cavities both with the same finesse of ~200, the lengths of which are ~430 mm and ~450 mm, respectively. Both cavities are locked to the resonance with the frequency of writing/reading laser. For more effectively filtering the photon noise of the incoherent components, we let the writing/reading laser beam doubly pass through $MC_1$ and $MC_2$.

The optical spectral filters (OSF) include 18 planar Fabry-Perot etalons. The free-spectral range and the finesse for each etalon are ~20 GHz and 6, respectively. The length of each etalon is stabilized to resonate with the signal light by a temperature controller. In the resonating case, its transmission is ~98% for the signal light and 16% for the writing/reading light. The total transmission of OSF is 66% for the signal light and ~$10^{-13}$ for the writing/reading light. After the OSF is used, the measured photon numbers of reading laser pulse entering each single-photon detector are reduced to $10^{-5}$ photons per pulse in the absence of the input signal light and the cold atoms.

The total detection efficiency $\eta_d \approx 19\%$, which is a combination of the pin-hole transmission (65%), the transmission of the set of optical filters (65%), the efficiency (90%) of fiber coupling to single-photon detectors $D_1$ or $D_2$ and the quantum efficiency of D1 and D2 (50%).

**Retrieval efficiencies for two storage systems with Zeeman degeneracy and lifting-Zeeman degeneracy**

The retrieval efficiency is defined as:

$$R_e(t) = \frac{\int \langle |\hat{\varepsilon}_{out}(t)|^2 \rangle dt}{\int \langle |\hat{\varepsilon}_{in}(t)|^2 \rangle dt} \qquad (1)$$

where $\int \langle |\hat{\varepsilon}_{in}(t)|^2 \rangle dt$ and $\int \langle |\hat{\varepsilon}_{out}(t)|^2 \rangle dt$ correspond to the photon numbers of the input and retrieval signal pulses, respectively. An arbitrarily-polarized signal light field can be expressed as a superposition of $\sigma^+$ – polarized and $\sigma^-$ – polarized components, which respectively couple to the $|a_m\rangle \leftrightarrow |e_{m+1}\rangle$ ( $m = -1, 0$ ) and $|a_m\rangle \leftrightarrow |e_{m-1}\rangle$ ( $m = 1, 0$ ) transitions for our experimental arrangement with the input signal field propagating along the quantum axis (z-direction). The writing/reading light is vertically-polarized, whose $\sigma^+$ – and $\sigma^-$ – polarized components respectively drive the $|b_m\rangle \leftrightarrow |e_{m+1}\rangle$ ( $m = -2, -1, 0$ ) and $|b_m\rangle \leftrightarrow |e_{m-1}\rangle$ ( $m = 2, 1, 0$ ) transitions.

Assuming that half of the cold atoms are prepared in state $|a_{m=1}\rangle$ and half in state $|a_{m=-1}\rangle$, we first discuss the retrieval efficiency of the $\sigma^+$ – polarized signal for the condition of a weak magnetic field and then for that of a moderate magnetic field. For the case of that the weak magnetic field is applied on the cold atoms, Zeeman sub-levels of the two

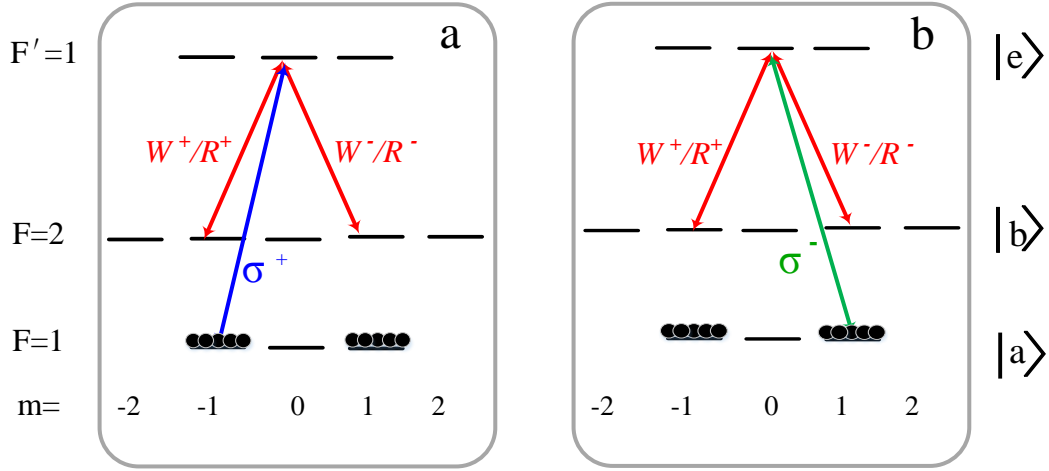

**Figure S1 Atomic level schemes for optical storage with the weak magnetic field ($B_0$=0.59G) and perfect atomic preparation.** (a) and (b) are for the storage of $\sigma^+-$ and $\sigma^--$polarized input signal photons, respectively.

ground F=1 and F=2 states are degenerate. In this case, the EIT system coupled by the $\sigma^+-$polarized input signal field is a typical tripod-type configuration including four levels $|a_{m=-1}\rangle - |b_{m=1}\rangle - |e_{m=0}\rangle - |b_{m=-1}\rangle$ (see Figure S1(a)). In such EIT system, the storage and retrieval processes for the signal field $\hat{\varepsilon}_{in}^+(z,t)$ can be described by the generalized dark-state polariton concepts [1,2]:

$$\hat{\Psi}^+(z,t) = \cos\vartheta\,\hat{\varepsilon}_{in}^+(z,t) - \sin\vartheta\sqrt{N}\left(\cos\Theta\,\hat{S}_{-1,-1}(z,t) + \sin\Theta\,\hat{S}_{-1,1}(z,t)\right) \quad (2)$$

where, the mixing angle is defined as $\tan\vartheta(t) = g\sqrt{N}/\sqrt{|\Omega_{C+}(t)|^2 + |\Omega_{C-}(t)|^2}$; $\Omega_{C+}(t)$ and $\Omega_{C-}(t)$ (C denotes W or R) are the Rabi frequencies of the right- and left-circularly-polarized writing ($W^+$ and $W^-$) or reading ($R^+$ and $R^-$) beam, respectively; $\cos\Theta = |\Omega_{C+}|/\sqrt{|\Omega_{C+}|^2 + |\Omega_{C-}|^2}$, $\sin\Theta = |\Omega_{C-}|/\sqrt{|\Omega_{C+}|^2 + |\Omega_{C-}|^2}$ ; $\hat{S}_{m,m'}(z,t)$ is the collective atomic spin operator, defined as $\hat{S}_{m,m'}(z,t) = (N_z)^{-1} \sum_{j=1}^{N_z} |a_m\rangle_{jj}\langle b_{m'}| e^{i\omega_{ab}t}$, $N_z(\gg 1)$ is the number of atoms in an interval $\Delta z$ [3]. By adiabatically switching off the writing beam ($\Omega_{C\pm}(t) \to 0$) over the small time interval $t_1$, $\hat{\varepsilon}_{in}^+(z,t)$ is mapped onto the superposition of the SWs $\hat{S}_{-1,-1}$ and $\hat{S}_{-1,1}$, which are expressed as [1, 2]:

$$\hat{S}_{-1,-1}(z,t_1) \propto \sqrt{\eta_W} \cos\Theta \hat{\varepsilon}_{in}^+(z-z_{01}, t=0), \qquad (3a)$$

$$\hat{S}_{-1,1}(z,t_1) \propto \sqrt{\eta_W} \sin\Theta \hat{\varepsilon}_{in}^+(z-z_{01}, t=0), \qquad (3b)$$

where $z_{01} = \int_0^{t_1} c\cos^2\vartheta(t)dt$, $\eta_W = \frac{N_e}{N_{in}}$ is the writing efficiency, $N_e$ and $N_{in}$ are the numbers of SW excitations and incoming photons at the time $t=0$, respectively. After storage time $t$, the SWs evolve into:

$$\hat{S}_{-1,-1}(z,t) \propto \sqrt{\eta_W} \cos\Theta \hat{\varepsilon}_{in}^+(z-z_{01}, 0) e^{-\frac{t}{2\tau_{-1,-1}} - i\omega_{-1,-1}t}, \qquad (4a)$$

$$\hat{S}_{-1,1}(z,t) \propto \sqrt{\eta_W} \sin\Theta \hat{\varepsilon}_{in}^+(z-z_{01}, 0) e^{-\frac{t}{2\tau_{-1,1}} - i\omega_{-1,1}t}, \qquad (4b)$$

where $\omega_{m,m'} = \frac{\mu_B B_0}{\hbar}(g_a(m+m') - \delta g m')$ and $\tau_{m,m'}$ are the Larmor frequency and the lifetime for the stored SW $\hat{S}_{m,m'}(z,t)$, respectively, the Landé factors $g_a \approx 0.4998$, $g_b \approx -0.5018$ and $\delta g = g_a + g_b \approx -0.002$. At time $t$, if the reading beam is switched on, the SWs will be mapped into the optical signal field $\hat{\varepsilon}_{out}^+(t)$ [1-2]:

$$\hat{\varepsilon}_{out}^{+}(z,t) \propto \sqrt{\eta_R}\left(\sin\Theta \hat{S}_{-1,1}(z,t) + \cos\Theta \hat{S}_{-1,-1}(z,t)\right)$$
$$\propto \sqrt{\eta_W \eta_R}\hat{\varepsilon}_P^{in}(z-z_0,0)(\sin^2\Theta e^{\frac{-t}{2\tau_{-1,1}}-i\omega_{-1,1}t} + \cos^2\Theta e^{\frac{-t}{2\tau_{-1,-1}}-i\omega_{-1,-1}t}), \quad (5)$$

where, $\eta_R = \frac{N_{out}}{N_e'}$ is the reading efficiency, $N_e'$ and $N_{out}$ are the numbers of SW excitations and retrieval photons at the time $t$, respectively. Combining Eq.(1), we can calculate the retrieval efficiency $R_e(t)$ of the $\sigma^+$-polarized signal photons in the EIT system $|a_{m=-1}\rangle - |b_{m=-1}\rangle - |e_{m=0}\rangle - |b_{m=1}\rangle$:

$$R_e^+(t) = R_{e0}\left|\sin^2\Theta e^{\frac{-t}{2\tau_1}-i\omega_{-1,1}t} + \cos^2\Theta e^{\frac{-t}{2\tau_2}-i\omega_{-1,-1}t}\right|^2 \quad (6)$$

where $R_{e0} = \eta_W \eta_R$, corresponding to the retrieval efficiency at the storage time of $t=0$. In the tripod-type EIT system $|a_{m=-1}\rangle - |b_{m=-1}\rangle - |e_{m=0}\rangle - |b_{m=1}\rangle$, we have $|\Omega_{C+}(t)| = |\Omega_{C-}(t)|$ and $\cos^2\Theta = \sin^2\Theta = 1/2$ according to the $^{87}$Rb data. Substituting $\cos^2\Theta = \sin^2\Theta = 1/2$ into Eq.(6), we express the retrieval efficiency as:

$$R_e^+(t) = \frac{1}{4}R_{e0}\left|e^{\frac{-t}{2\tau_{-1,1}}-i\omega_{-1,1}t} + e^{\frac{-t}{2\tau_{-1,-1}}-i\omega_{-1,-1}t}\right|^2. \quad (7)$$

The above analyses for obtaining the retrieval efficiency of $\sigma^+$-polarized signal field are available for that of the $\sigma^-$-polarized one since the EIT atomic-level systems used for storing the two signals are totally symmetric (see Fig.S1(a) and (b)). The retrieval efficiency of the $\sigma^-$-polarized signal photons is

$$R_e^-(t) = \frac{1}{4} R_{e0} \left| e^{\frac{-t}{2\tau_{1,-1}} - i\omega_{1,-1}t} + e^{\frac{-t}{2\tau_{1,1}} - i\omega_{1,1}t} \right|^2 \qquad (8)$$

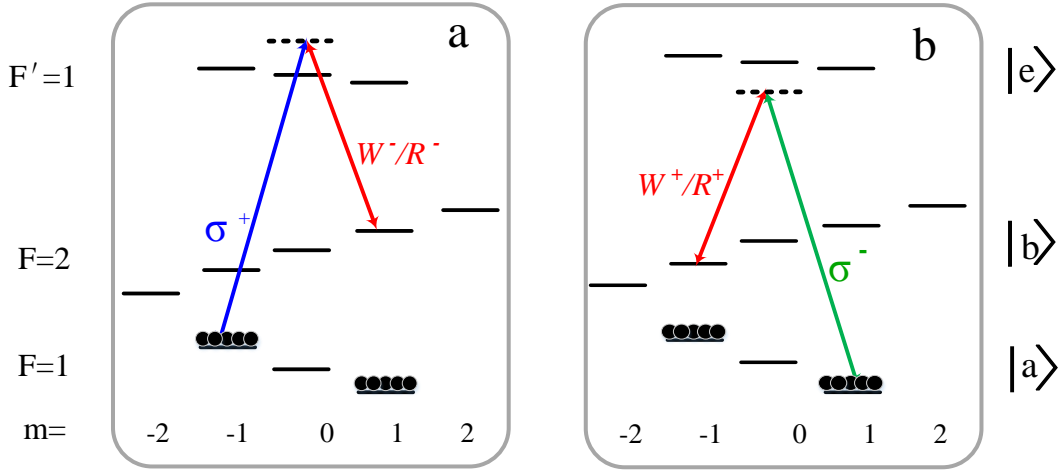

**Figure S2. Atomic level schemes for optical storage with the moderate magnetic field ($B_0$=13.5G) and perfect atomic preparation.** (a) and (b) are for the storage of $\sigma^+-$ and $\sigma^- -$polarized input signal photons, respectively.

When a moderate magnetic field is applied on the cold atoms, the Zeeman sublevels will be lifted and in this case the magnetic-field-sensitive transition $|a_{m=-1}\rangle \leftrightarrow |b_{m=-1}\rangle$ will not match the EIT-two-photon resonance, while the magnetic-field-insensitive transition $|a_{m=-1}\rangle \leftrightarrow |b_{m=1}\rangle$ will. Thus, the four-level tripod EIT system is changed into the three-level Λ-type EIT system ($|a_{m=-1}\rangle - |e_{m=0}\rangle - |b_{m=1}\rangle$) (Fig.S2(a)). The storage process in such Λ-type EIT system can be described by the dark-state porlariton concept [3], $\hat{\Psi}^+(z,t) = \cos\vartheta \hat{\varepsilon}_{in}^+(z,t) - \sin\vartheta\sqrt{N}\hat{S}_{-1,1}(z,t)$, where the mixing angles is defined as $\tan\vartheta(t) = g\sqrt{N}/|\Omega_{C+}(t)|$. The $\sigma^+$-polarized signal light field $\hat{\varepsilon}_{in}^+(z,t)$ will be only stored as the magnetic-field-insensitive SW $\hat{S}_{-1,1}(z,t)$, which is expressed as [3]:

$$\hat{S}_{-1,1}(z,t_1) \propto \sqrt{\eta_W}\hat{\varepsilon}_{in}^+(z-z_{01},0), \qquad (9)$$

Using a analysis similar to obtain Eq.(7), we can calculate the retrieval efficiency of the $\sigma^+$-polarized signal photons for the case of Zeeman degeneracy:

$$R_e^+(t) = R_{e0}\left|e^{\frac{-t}{2\tau_{-1,1}} - i\omega_{-1,1}t}\right|^2 = R_{e0}e^{-\frac{t}{\tau_{-1,1}}} \qquad (10)$$

The above analyses for obtaining the expression of the retrieval efficiency of $\sigma^+$-polarized signal field are available for that of the $\sigma^-$-polarized one since the EIT atomic-level systems used for storing the two signals are totally symmetric (see Fig.S2(a) and (b)). The retrieval efficiency of the $\sigma^-$-polarized signal photons is

$$R_e^-(t) = R_{e0}e^{-\frac{t}{\tau_{1,-1}}}. \qquad (11)$$

The magnetic-field-insensitive SWs $\hat{S}_{-1,1}(z,t)$ and $\hat{S}_{1,-1}(z,t)$ should have the same lifetime, i.e., $\tau_{-1,1} = \tau_{1,-1} = \tau_1$, so both retrieval efficiencies for the $\sigma^+$- and $\sigma^-$-polarized signal for the case of lifting Zeeman degeneracy can be expressed as:

$$R_e(t) = R_{e0} e^{-\frac{t}{\tau_1}} \qquad (12)$$

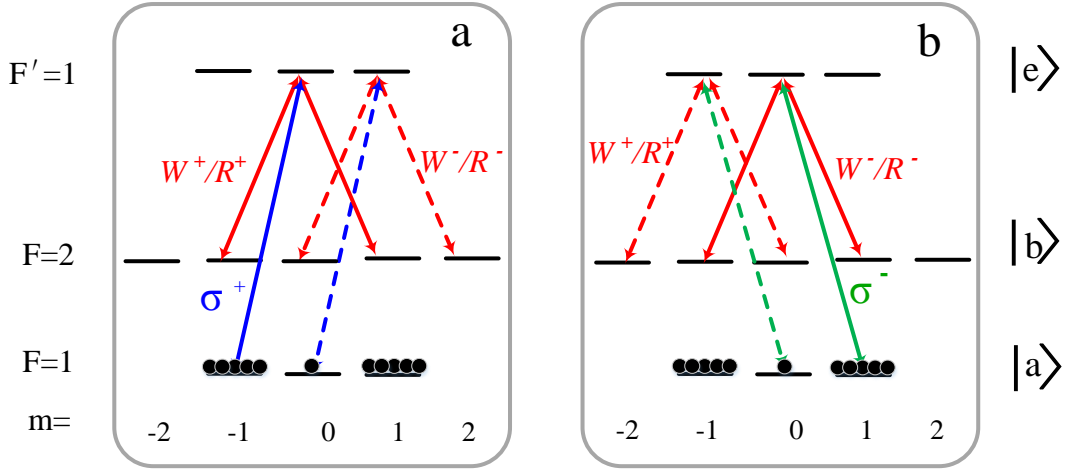

**Figure S3. Atomic level schemes for optical storage with the weak magnetic field ($B_0$=0.59G) and imperfect atomic preparation**. (a) and (b) are for the storage $\sigma^+-$ and $\sigma^--$polarized input signal photons, respectively.

In our practical experiment, the atomic preparation is not perfect when the weak magnetic field ($B_0$=0.59G) is applied on the atoms. In this case, the paucity of atoms is retained in state $|a_{m=0}\rangle$ after the optical pumping, and thus another tripod-type EIT system including the four levels $|a_{m=0}\rangle - |b_{m=0}\rangle - |e_{m=1}\rangle - |b_{m=2}\rangle$ (See Fig.S3(a)) is formed when the input signal photons are $\sigma^+$-polarized. So its contribution to the storage of $\sigma^+$-polarized signal photons should be considered also. Using a analysis similar to obtain Eq.(6), we can write the retrieval efficiency of the $\sigma^+$-polarized signal in such a tripod-type EIT system. Combining the contributions from the two independent tripod-type EIT systems $|a_{m=-1}\rangle - |b_{m=-1}\rangle - |e_{m=0}\rangle - |b_{m=1}\rangle$ and $|a_{m=0}\rangle - |b_{m=0}\rangle - |e_{m=1}\rangle - |b_{m=2}\rangle$, the total retrieval efficiency of the $\sigma^+$-polarized signal light for the case with Zeeman degenerate levels can be written as:

$$R_e^+(t) = R_{e0} \left| p_1^+ \left( \frac{1}{2} e^{-t/2\tau_1 - i\omega_{1,1}t} + \frac{1}{2} e^{-t/2\tau_{-1,-1} - i\omega_{1,-1}t} \right) + p_0^+ \left( \frac{1}{7} e^{-t/2\tau_{0,0} - i\omega_{0,0}t} + \frac{6}{7} e^{-t/2\tau_{0\,2} - i\omega_{0,2}t} \right) \right|^2 \quad (14)$$

where $p_1^+$ and $p_0^+$ are the storage weight of the input signal in the two tripod EIT systems, respectively, which mainly dependent on atomic population in the state $|a_{m=-1}\rangle$ and $|a_{m=0}\rangle$, respectively.

All the above analyses for obtaining the retrieval efficiency of $\sigma^+$-polarized signal field are available for that of the $\sigma^-$-polarized one since the EIT atomic-level systems $|a_{m=1}\rangle - |b_{m=1}\rangle - |e_{m=0}\rangle - |b_{m=-1}\rangle$ and $|a_{m=0}\rangle - |b_{m=0}\rangle - |e_{m=-1}\rangle - |b_{m=-2}\rangle$ used for storing the two signals are totally symmetric (see Fig.S3(a) and (b)). The retrieval efficiency of $\sigma^-$-polarized signal for the case of Zeeman degeneracy can be written as:

$$R_e^-(t) = R_{e0} \left| p_1^- \left( \frac{1}{2} e^{-t/2\tau_1 - i\omega_{1,-1}t} + \frac{1}{2} e^{-t/2\tau_{1,1} - i\omega_{1,1}t} \right) + p_0^- \left( \frac{1}{7} e^{-t/2\tau_{0,0} - i\omega_{0,0}t} + \frac{6}{7} e^{-t/2\tau_{0\,2} - i\omega_{0,-2}t} \right) \right|^2 \quad (15)$$

where $p_1^-$ and $p_0^-$ are the storage weight of the input signal in the two tripod EIT systems, respectively, which mainly dependent on atomic population in the state $|a_{m=1}\rangle$ and $|a_{m=0}\rangle$, respectively.

Based on Eq.(14) and Eq.(15), we give the fittings (the green lines) to the experimental data (red circular points) in Fig.2(a) and (b), respectively, yielding the best-fit values of $p_1^\pm = 0.93$, $p_0^\pm = 0.07$, $\tau_{0,0} = 2\ ms$, $\tau_1 = 1.6\ ms$, $\tau_{1,1} = \tau_{-1,-1} = \tau_{0,2} = \tau_{0,-2} = 30\ \mu s$.

## Quantum Tomography

The quantum process tomography requires to implement an experimental reconstruction of the quantum tomography matrix $\chi$, which maps an input state $\rho_{in}$ onto the corresponding output state $\rho_{out}$ via:

$$\rho_{out} = \sum_{m,n} \chi_{m,n} \hat{\sigma}_m \rho_{in} \sigma_n^\dagger$$

where $\hat{\sigma}_i$ are the Pauli spin operators ($i=m,n$), $\dagger$ denotes the adjoint operator. The quantum process fidelity is defined as $F_{process} = Tr\left(\sqrt{\sqrt{\chi}\chi_{ideal}\sqrt{\chi}}\right)^2$ with $\chi_{ideal} = \begin{pmatrix} 1 & 0 & 0 & 0 \\ 0 & 0 & 0 & 0 \\ 0 & 0 & 0 & 0 \\ 0 & 0 & 0 & 0 \end{pmatrix}$. For reconstructing the quantum process $\chi$, we carry out the storage and retrieval for four input signal states: $|H\rangle$, $|V\rangle$, $|D\rangle$ and $|R\rangle$. For each input state, a quantum state tomography is used to reconstruct its output state $\rho_{out}$. To implement the reconstruction, the retrieved photon qubit is measured in three bases $H-V$, $D-A$ and $R-L$. For obtaining a positive Hermitian and trace preserving $\chi$, Maximum Likelihood Estimation [4] is applied. The errors of process fidelity are calculated via Monte Carlo Simulation [4].

## Evaluation of quantum process fidelity

The quantum process fidelity of retrieved signal photons will decrease with the storage time due to the decoherence mechanism of SWs. With the experimental data of the quantum process matrix $\chi$, one can calculate the quantum process fidelity for any storage time according to the definition $F_{Process} = Tr\left(\sqrt{\sqrt{\chi}\chi_{ideal}\sqrt{\chi}}\right)^2$. However, from this definition, the dependence of the quantum process fidelity $F_P$ on the decoherence of SWs is not very obvious. It is significant to establish an expression of $F_{process}$ depending on the storage time and involving the decoherence of the SWs. The polarization of input signal photons can be characterized by the Stokes parameters $S_i$, thus we will deduce the expression starting from calculating $S_i$ of the input and the retrieval (output) signal light fields.

The input signal field $\hat{\varepsilon}_{in}(t)$ with arbitrary photonic polarization can be regarded as the superposition of $|R\rangle$ and $|L\rangle$ components and expressed by:

$$\hat{\varepsilon}_{in}(t) = E_0(t)\left(\alpha|R\rangle + \beta e^{i\theta}|L\rangle\right) \qquad (16)$$

Here $|R\rangle$ and $|L\rangle$ correspond to the right circular ($\sigma^+$) polarization and left circular ($\sigma^-$) polarization, respectively. $\alpha$ and $\beta$ are normalized amplitudes, $\alpha^2 + \beta^2 = 1$. $\theta$ is the relative phase between the two components. $E_0(t)$ is slowly varying envelope. So, the Stokes parameters [5] of the input signal field can be written as:

$$\begin{pmatrix} S_0^{in} \\ S_1^{in} \\ S_2^{in} \\ S_3^{in} \end{pmatrix} = I_0 \begin{pmatrix} 1 \\ 2\alpha\beta \sin\theta \\ \alpha^2 - \beta^2 \\ 2\alpha\beta \cos\theta \end{pmatrix} \qquad (17)$$

where $I_0 \propto |E_0|^2$ is the intensity of the input signal light, $\alpha = \beta = 1/\sqrt{2}$, $\theta = 0, \pi, \pi/2, 3\pi/2$ correspond to $|H\rangle$, $|V\rangle$, $|D\rangle$, $|A\rangle$ input polarization states, as well as $\{\alpha=1, \beta=0 \text{ or } \alpha=0, \beta=1\}$ to $\{|R\rangle$ or $|L\rangle\}$ input polarization states. At time $t=0$, the optical signal field $\hat{\varepsilon}_{in}(t) = E_0(t)(\alpha|R\rangle + \beta e^{i\theta}|L\rangle)$ is mapped onto the superposition $\hat{\tilde{S}}(z,0) \propto \alpha S_{1,-1}(z,0) + \beta e^{i\theta} S_{-1,1}(z,0)$ and stored in the cold atoms cloud. After a time delay $t$, the stored SW evolves into $\hat{\tilde{S}}(z,t) = e^{-t/\tau_1}\alpha S_{1,-1}(z,t) + e^{-t/\tau_1}\beta e^{i\theta+i\delta\phi+i\phi} S_{-1,1}(z,t)$, where $\phi = (\omega_{1,-1} - \omega_{-1,1})t = 2\mu_B B_0(g_a + g_b)t/\hbar$ is the relative phase between the two SWs resulting from the Larmor precession in the bias magnetic field $B_0$ and $\delta\phi = 2\mu_B(g_a+g_b)/\hbar \int_0^t \delta B dt$ is the fluctuation of $\phi$ resulting from the magnetic-field fluctuation $\delta B$, $g_a$ and $g_b$ are Landé factors. We then transfer the SW $\hat{S}(z,t)$ into photon pulse at time $t$. The Stokes parameters of the retrieval photons can be written as:

$$\begin{pmatrix} S_0^{out} \\ S_1^{out} \\ S_2^{out} \\ S_3^{out} \end{pmatrix} = \eta_d R_e(t) I_0 \begin{pmatrix} 1 \\ 2\alpha\beta \sin(\theta+\delta\phi) \\ \alpha^2 - \beta^2 \\ 2\alpha\beta \cos(\theta+\delta\phi) \end{pmatrix} + \eta_d \begin{pmatrix} 2I_N \\ 0 \\ 0 \\ 0 \end{pmatrix} \approx \eta_d \begin{pmatrix} I_0 R_e(t) + 2I_N \\ 2I_0 R_e(t)\alpha\beta\gamma(t)\sin\theta \\ I_0 R_e(t)(\alpha^2-\beta^2) \\ 2I_0 R_e(t)\alpha\beta\gamma(t)\cos\theta \end{pmatrix} \qquad (18)$$

where, $\eta_d$ is the total detection efficiency, $I_N$ is the intensity of the background noise in the signal channel. The noise includes spontaneous emissions from the atoms, stray light and leakage of the reading beam and is assumed to be unpolarized. The spin-wave lifetime $\tau_1$ is absorbed by retrieval efficiency $R_e(t)$. Both the retrieval efficiencies for $|R\rangle$ and $|L\rangle$ components of the input signal have been assumed as $R_e(t)$ and the relative phase ($\phi$) between the two components has been perfectly compensated. $\gamma(t) = \langle \cos\delta\phi \rangle = e^{-t^2/\sigma_\gamma^2}$ is dephasing factor, which has a Gaussian distribution [5] with $1/e$ dephasing time $\sigma_\gamma$. The root-mean-square (rms) width $\sigma_B$ of the magnetic-field fluctuation can be obtained by [5] $\sigma_B = \dfrac{1}{\sigma_\gamma} \dfrac{1}{\sqrt{2}(g_a+g_b)} \dfrac{\hbar}{\mu_B}$ .

The quantum state fidelity is defined by

$$F_{st} = Tr(\rho_{in}\rho_{out}) \ . \tag{19}$$

The input (output) density matrices can be calculated by $\rho_{in(out)} = \dfrac{1}{2}\sum_{i=0}^{3}\dfrac{S_i^{in(out)}}{S_0}\hat{\sigma}_i$ , where $\hat{\sigma}_i$ are the Pauli spin operator. From the Eqs (18), and (19), we can calculate the quantum state fidelities for the six input states: $|H\rangle = (|R\rangle+|L\rangle)/\sqrt{2}$, $|V\rangle = (|R\rangle-|L\rangle)/\sqrt{2}$, $|A\rangle = (|R\rangle-i|L\rangle)/\sqrt{2}$, $|D\rangle = (|R\rangle+i|L\rangle)/\sqrt{2}$, $|R\rangle$ and $|L\rangle$, which are

$$F_{st(X)} \approx \frac{(1+e^{-t^2/\sigma_\gamma^2})\eta_d R_e(t) + 2N}{2(\eta_d R_e(t) + 2N)},$$

for $|X\rangle = |H\rangle, |V\rangle, |A\rangle, |D\rangle$ and

$$F_{st(X)} \approx \frac{\eta_d R_e(t) + N}{\eta_d R_e(t) + 2N},$$

for $|X\rangle = |R\rangle, |L\rangle$, where $N = \frac{\eta_d I_N}{I_0}$ is the relative background noise. For our experimental case of the input signal pulse with mean-photon number of $\bar{n} = \frac{I_0 \Delta\tau}{\hbar\omega} \approx 1$, $N \approx \eta_d \frac{I_N \Delta\tau}{\hbar\omega}$, which can be obtained by recording counts at each single-photon detector during the reading window of $\Delta\tau$ when the reading beam is sent while input signal light is blocked. The quantum process fidelity can be expressed as [6]

$$F_{process} = \frac{3\overline{F_{st}} - 1}{2} \tag{20}$$

where $\overline{F_{st}} = (F_{st(H)} + F_{st(V)} + F_{st(D)} + F_{st(A)} + F_{st(L)} + F_{st(R)})/6$ is the average quantum state fidelity. Finally, the expression of the quantum process fidelity is denoted as:

$$F_{process} \approx \frac{(1+e^{-t^2/\sigma_\gamma^2})\eta_d R_e(t) + N}{2(\eta_d R_e(t) + 2N)} \tag{21}$$